\documentclass{iaus} 
\usepackage{graphicx}

\title[CRTS] 
{The Catalina Real-time Transient Survey}

\author[Drake et al.]   
{A.J. Drake$^1$, S.G. Djorgovski$^{1,6}$, A. Mahabal$^1$, J.L. Prieto$^2$,\\ 
E. Beshore$^4$, M.J. Graham$^1$,  M. Catalan$^3$, S. Larson$^4$, \\
E. Christensen$^5$, C. Donalek$^1$ \and R. Williams$^1$
}

\affiliation{$^1$California Institute of Technology, Pasadena, CA 91125, USA. 
\\ email: {\tt ajd@cacr.caltech.edu} \\[\affilskip]
$^2$Dept. of Astrophysical Scinces, Princeton University, NJ 08544, USA.\\[\affilskip] 
$^3$Depto. de Astronomia y Astrofisica, Pont. Uni. Catolica de Chile, Santiago, Chile.\\[\affilskip]
$^4$Lunar and Planetary Lab, University of Arizona, Tucson, AZ 85721, USA.\\[\affilskip] 
$^5$Gemini South Observatory, c/o AURA, Casilla 603, La Serena, Chile.\\[\affilskip]
$^6$Distinguished Visiting Professor, King Abdulaziz Univ., Jeddah, Saudi Arabia.
}

\pubyear{2011}
\volume{285}  
\pagerange{1}
\jname{New Horizons in Time Domain Astronomy}
\editors{E. Griffin, R. Hanisch \& R. Seaman, eds.}
\begin{document}

\maketitle

\begin{abstract}

  The Catalina Real-time Transient Survey (CRTS) currently covers 33,000 deg$^2$ of the sky in search of transient
  astrophysical events, with time baselines ranging from 10 minutes to $\sim 7$ years. Data provided by the Catalina Sky
  Survey provides an unequaled baseline against which $> 4,000$ unique optical transient events have been discovered and
  openly published in real-time. Here we highlight some of the discoveries of CRTS.

\end{abstract}


\firstsection 
\section{Introduction}

For the past four years the Catalina Real-time Transient Survey (CRTS; Drake et al.~2009a, Djorgovski et al. 2011,
Mahabal et al. 2011) has systematically surveyed tens of thousands of square degrees of the sky for transient
astrophysical events.  CRTS discovers highly variable and transient objects in real-time, making all discoveries public
immediately, thus benefiting a broad astronomical community. Data is leveraged from three telescopes used in a search
for NEOs, operated by LPL, which cover up to $\sim$ 2,500 deg$^2$ per night with four exposures separated by $\sim$ 10
mins.  The total survey area is $\sim$ 33,000 deg$^2$ and reaches depth V $\sim$ 19 to 21.5 mag (depending on telescope)
during 23 nights per lunation.  All data are automatically processed as they are taken, and optical transients (OTs) are
immediately distributed using a variety of electronic mechanisms (see http://www.skyalert.org/, and
http://crts.caltech.edu/). CRTS has so far discovered $>$ 4,000 unique OTs including $> 1,000$ supernovae and 500 dwarf
novae.

\section{Discoveries}

{\underline{\it Supernovae and their hosts}}.  Supernovae are both cosmological tools and probes of the final states of
stellar evolution.  While many astronomical surveys focus on type Ia SNe, being standard candles, CRTS uses its wide
area coverage to look for rare types of events that may be missed by many traditional SN surveys. With $> 1,000$ SNe (CRTS
published more SN discoveries in both 2009 and 2010 than any other survey), this data set has allowed us to carry
out a systematic exploration of supernova properties leading to the discovery of extremely luminous supernovae and
supernovae in extremely faint host galaxies, with $M_V \sim -12$ to $-13$, i.e., $\sim 0.1$\% of $L_*$.

Two especially interesting classes of luminous SNe discovered by CRTS, TSS and PTF include luminous type-Ic SNe (SN
2005ap, Quimby et al.~2007; SN 2009de, Drake et al.~2009b, 2010; SN 2009jh, Drake et al.~2009c, Quimby et al.~2011; SN
2010gx, Mahabal et al.~2010, Pastorello et al.~2010a, 2010b; and CSS110406:135058+261642, Drake et al.~2011b) and
ultra-luminous and energetic type-IIn SNe (SN 2008fz, SN 2009jg, etc., Drake et al.~2009c,2010,2011a). These supernovae
have been found to favor extremely faint-host galaxies (Drake et al.~2009a, 2010) suggesting the importance of
host-galaxy environment and explaining why more such events have not been discovered previously.  In Figure 1, we
contrast the SN host-galaxy absolute magnitudes from CRTS, with those from the long running Lick Observatory SN Search
(LOSS; Filippenko et al.~2001) which concentrates on bright nearby galaxies.

The rate of our SN discovery in intrinsically faint galaxies implies phenomenally high specific SN rates (Drake et
al.~2009a).  Although such galaxies are common, a very small fraction of all baryonic matter is expected in them
(Kauffmann et al.~2003).  Evidence suggests that these galaxies include blue compact dwarfs and irregular dwarfs, where
excessive star formation rates accelerate SNe rates for the most rapidly evolving massive stars (progenitors of luminous
SN).  Additional evidence for enhancements in the rates of SN-Ia, up to 1500\%, has been speculated by Della Valle \&
Panagia (2003).

\begin{figure}[b]
\begin{center}
 \includegraphics[width=2.4in]{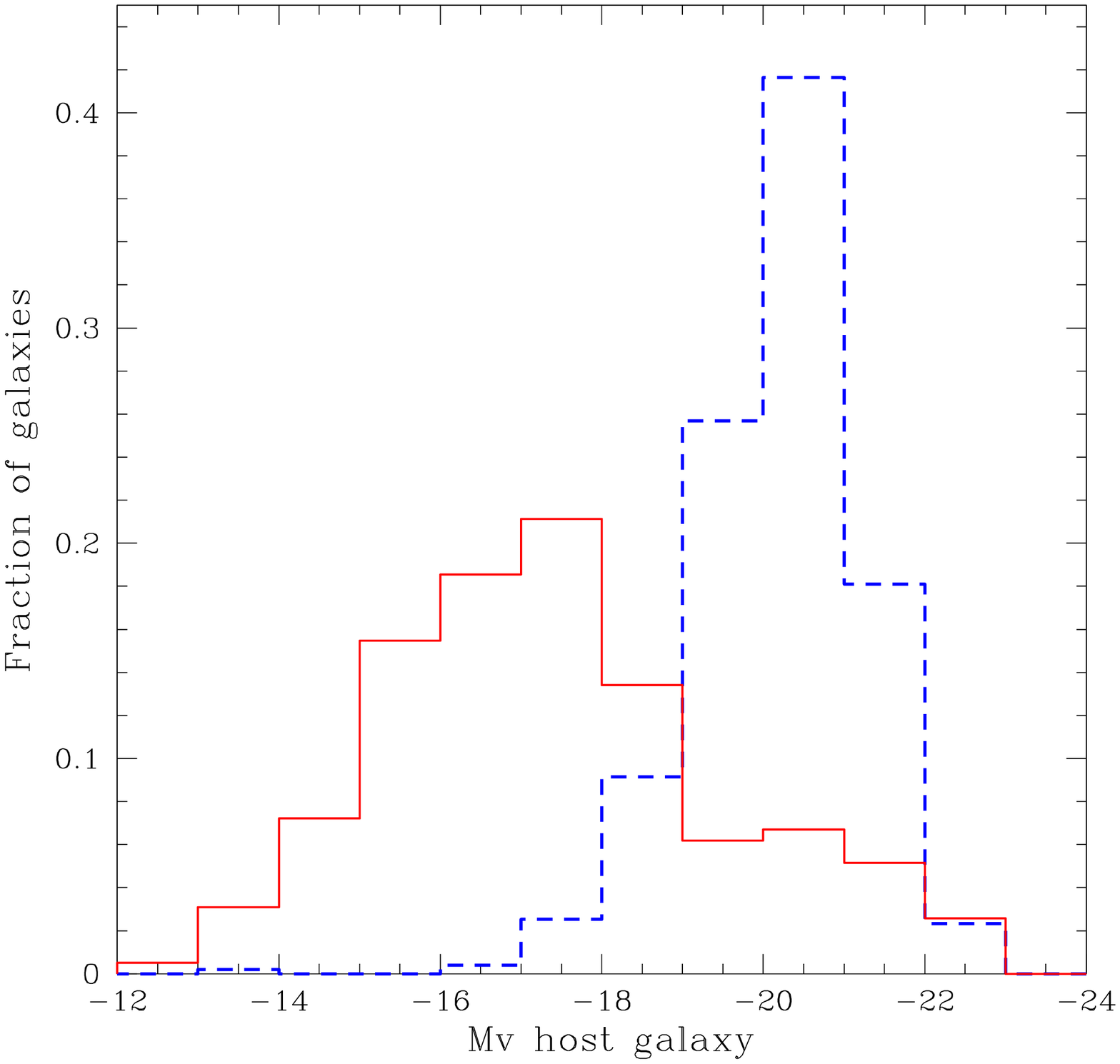} \includegraphics[width=2.4in]{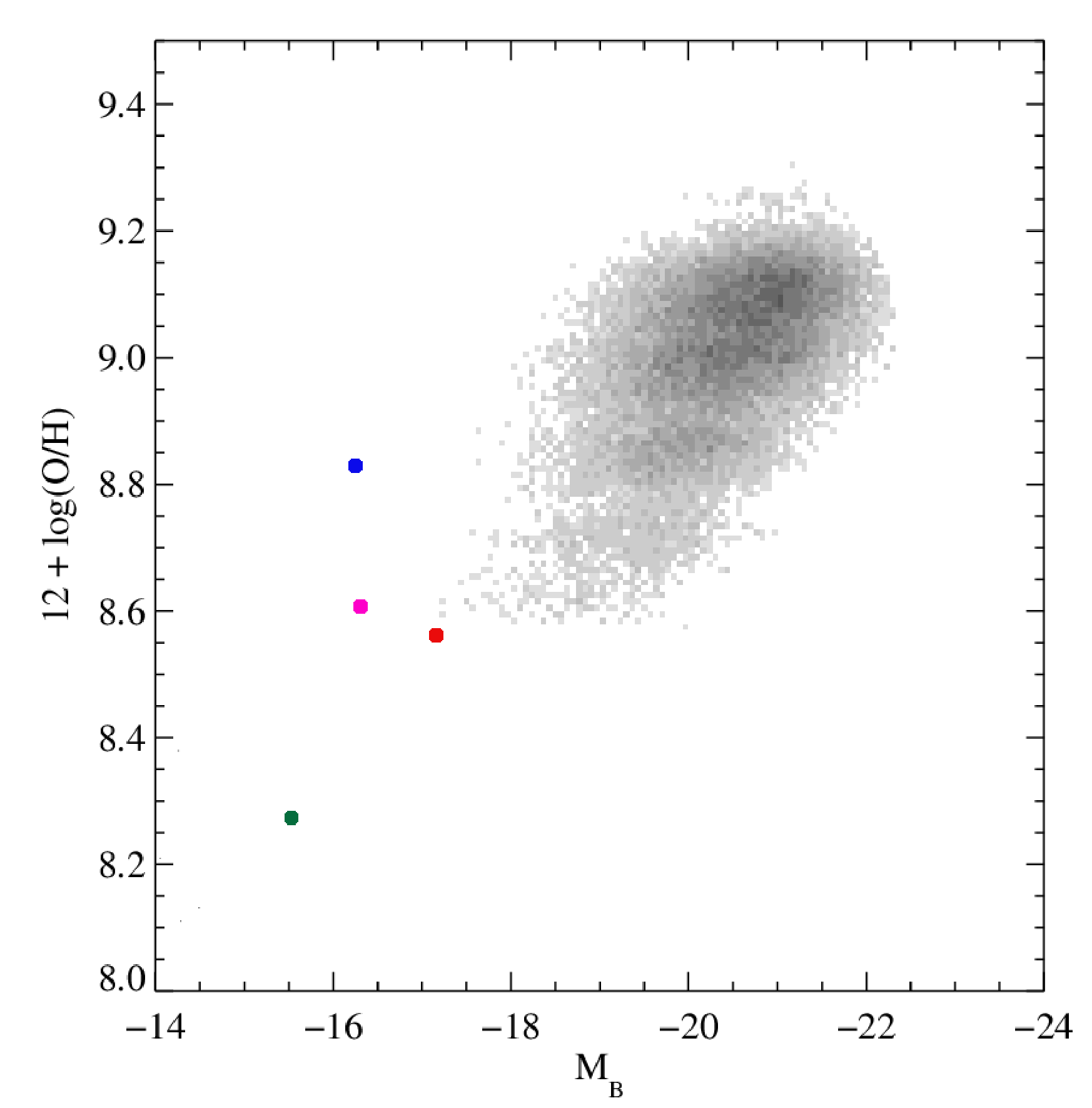} 
 \caption{SN Hosts.
Left: A comparison of supernova host-galaxy absolute magnitudes. Solid line: CRTS. Short-dashed line the 
Lick Observatory Supernova Search (LOSS). Right: Host luminosity and metalicity for four energetic type 
IIn SN host galaxies, compared with 53,000 star-forming galaxies from SDSS (Tremonti et al. 2004).
}
\label{fig1}
\end{center}
\end{figure}

It is likely that these dwarf galaxies have low metallicities due to a delayed onset of star formation and expulsion of
enriched SN ejecta from their shallow potential wells.  Based on the galaxy mass-metalicity relationship (Tremonti et
al.~2004), low-luminosity hosts are expected to be low-metalicity hosts.  This prediction was recently confirmed in the
work of by Neill et al.~(2011), Stoll et al.~(2011) and Kozlowski et al.~(2010) as well as in our recent work shown in
Figure 1.  Low metallicities are speculated to lead to a top-heavy IMF, which would account both for an enhanced
specific SN rate, and the propensity for highly luminous events (from high-mass progenitors). Low metalicity host 
galaxies are also linked to the broad-line type-Ic hypernovae associated with long-timescale GRBs (Stanek et al.~2006).

Another interesting discovery is of a new class of SNe that may be associated with AGN
accretion disks.  The likely the most luminous and optically energetic SN ever discovered, CSS100217, within the AGN
disk of a bright NLS1 galaxy, demonstrates that extreme supernova can occur in a variety of extreme environments (Drake
et al. 2011a).

{\underline{\it Blazars}}.
Blazars are highly variable optical and radio sources.  They are often targeted for optical
follow-up after their outbursts at other wavelengths. CRTS provides an unbiased, statistical optical monitoring of known blazar
sources over three quarters of the sky.  Due to the erratic nature of blazar variability and the association of these
sources with previously cataloged, and often faint radio sources, we have found several tens of likely blazars based on
transient outburst events.  We have also to produce variability-selected blazar counterparts to the previously
unidentified $Fermi$ $\gamma$-ray sources.  CRTS data are also being combined with
radio data from the Owens Valley Radio Observatory and will be used to provide better constraints to the theoretical
models of blazar emission and variability.

{\underline{\it Dwarf Novae and UV Ceti variables}}.
The CRTS project has discovered more than 500 new dwarf nova type cataclysmic variables (CVs). Since these objects are
found in real-time, the outbursts are often followed.  Thus far, 132 CV discoveries have been alerted to users of the VSNET
system (www.kusastro.kyoto-u.ac.jp/vsnet/), resulting in successful period determination in dozens of these systems.
Similarly, CRTS has discovered over 100 UV Ceti variables (flare stars) varying by several magnitudes within minutes.
The rate of such flares is still poorly constrained and must be understood so that future surveys can find rare types of
rapid transients. The short cadence of CRTS is well tuned to the discovery of these events. Another class of rapid
transients are eclipses of white dwarf binary systems. These systems probe the end state of stellar binary evolution. 
Although first discovered in real-time, archival searches revealed dozens more eclipsing systems, including some with
low mass companions (Drake et al. 2011c).

{\underline{\it Acknowledgments}}.  CRTS is supported by the NSF grant AST-0909182.  We thank the personnel of 
many observatories involved in the survey and the follow-up observations.

\end{document}